\documentclass[10pt,conference,letterpaper]{IEEEtran}

\usepackage[ruled,vlined,boxed]{algorithm2e}
\usepackage{graphicx}
\usepackage{url}
\usepackage{paralist}
\usepackage{algorithmic}
\usepackage{color}
\usepackage{amssymb, amsmath, latexsym}
\usepackage{comment}
\usepackage{subfigure}
\usepackage{balance}
\usepackage{etoolbox}

\newcommand{\Rho}{\mathrm{P}}

\newcommand{\ndnname}[1]{{\small \tt #1}}
\newcommand{\descr}[1]{\medskip\noindent\textbf{#1}}
\renewcommand{\paragraph}{\descr}

\begin{document}
\IEEEoverridecommandlockouts 

\title{Poseidon: Mitigating Interest Flooding DDoS Attacks in Named Data Networking}

\author{\IEEEauthorblockN{
Alberto Compagno\IEEEauthorrefmark{1},
Mauro Conti \IEEEauthorrefmark{1}\IEEEauthorrefmark{3}$^{,\#,+}$,
Paolo Gasti \IEEEauthorrefmark{2}\IEEEauthorrefmark{3}$^{,\#}$\thanks{This work was conducted in the context of the NSF project: CNS- 1040802: FIA: Collaborative Research: Named Data Networking (NDN).}\thanks{$^{\#}$Work performed in part while at UC Irvine. }\thanks{$^{+}$Mauro Conti was supported by Marie Curie Fellowship PCIG11-GA-2012-321980, funded by the European Commission for the PRISM-CODE project, and by the University of Padua Researchers' Mobility grant 2012. This work has been partially supported by the TENACE PRIN Project 20103P34XC funded by the Italian MIUR.},
Gene Tsudik \IEEEauthorrefmark{3} 
}
\IEEEauthorrefmark{1}University of Padua, Padua, Italy email: acompagn@studenti.math.unipd.it, conti@math.unipd.it
\\
\IEEEauthorrefmark{2}New York Institute of Technology,  New York, NY, USA - email: pgasti@nyit.edu
\\
\IEEEauthorrefmark{3}University of California, Irvine, CA, USA - email: gts@ics.uci.edu
}

\maketitle

\begin{abstract}
Content-Centric Networking (CCN) is an emerging networking paradigm being considered as a possible replacement for the current IP-based host-centric Internet infrastructure. 
CCN focuses on content distribution, which 
is arguably not well served by IP.
Named-Data Networking (NDN) is an example of CCN. NDN is also an active research project under the NSF Future Internet Architectures (FIA) program. FIA emphasizes security and privacy from the outset and by design. To be a viable Internet architecture, NDN must be resilient against current and emerging threats.

This paper focuses on distributed denial-of-service (DDoS) attacks; in particular we address {\em interest flooding}, an attack that exploits key architectural features of NDN. We show that an adversary with limited resources can implement such attack, having a significant impact on network performance. We then introduce Poseidon: a framework for detecting and mitigating interest flooding attacks. Finally, we report on results of extensive simulations assessing proposed countermeasure.
\end{abstract}

\section{Introduction}
\label{introduction}

The Internet is an amazing success story, connecting hundreds of millions of users.
The way people access and utilize it has changed radically since the 1970-s when its 
architecture was conceived. Today, the Internet has to accommodate new services, 
new usage models and new access technologies. Users are increasingly mobile,
constantly accessing -- and contributing to -- remote information using a variety
of devices such as laptops and smartphones. Ever-increasing mobility, device 
heterogeneity, as well as massive amounts of user-generated content and social networking 
are exposing the limits of the current Internet architecture. 

To this end, there are some recent research 
efforts~\cite{NDN,XIA,mobility-first,nebula,choicenet} with the long-term goal of 
designing and deploying a next-generation Internet architecture. 
One such new architecture is Named Data Networking (NDN). It is based on the principle 
of  Content-Centric Networking, where content -- rather than hosts -- occupies the central 
role 
in the communication architecture. NDN is one of the five NSF-sponsored Future Internet 
Architectures (FIA)~\cite{FIA}; like the rest, it is an on-going research effort.
NDN is primarily oriented towards efficient large-scale content distribution. Rather than 
establishing 
direct IP connections with a host serving content, NDN consumers directly request (i.e., 
express 
{\em interest} in) pieces of content by name; %
the network is in charge of finding the closest copy of the content, and of retrieving it 
as efficiently as possible. This decoupling of content and location allows NDN to 
efficiently implement multicast, content replication and fault tolerance.
One of the key goals of the NDN project is {\em ``security by design''}. In contrast 
to today's Internet, where security problems were (and are still being) identified along 
the way, the NSF FIA program (for all of its projects) stresses both awareness of 
issues and support for features and countermeasures from the outset. 
To this end, this paper investigates distributed denial of service (DDoS) attacks in NDN.
DDoS attacks are considered to pose serious threats to the current Internet. 
NDN is not immune to them and might actually offer avenues for new DDoS
attacks. 

In NDN, each content consumer's request (called an ``interest'') causes NDN 
routers to store a small amount of transient state, which is flushed as 
the content is routed back to the consumer. This has been pointed out 
in previous work as a plausible attack vector -- under the name of interest flooding 
attack \cite{ndn-dos,WahlishSV12}.
Motivated by the importance of addressing security in the early stages of a potential
new Internet architecture, we focus on DDoS over NDN, specifically, using interest flooding attack. 
We believe that interest flooding attack and countermeasures deserve an in-depth investigation before 
NDN can be considered ready for large-scale deployment.
While some preliminary results of this research appeared in \cite{NDN-ACSAC}, in the current paper  
we show, via extensive simulations,  that interest flooding attacks
are not just theoretical.  It is, in fact, relatively easy to perform interest flooding with 
rather limited resources. We simulate interest flooding over 
a realistic topology~\cite{Heckmann03onrealistic}: the AT\&T network. 
We focus on reactive countermeasures and propose techniques for early detection
of interest flooding. (This was left as an open problem in~\cite{ndn-dos}.) We then describe the 
design and implementation of Poseidon -- a framework for local and distributed interest flooding attack 
mitigation. Finally, we report on the effectiveness of proposed methods.

\paragraph{Organization.}
We present NDN in Section~\ref{sec:overview} and interest flooding in Section~\ref{approaches}.
Section~\ref{exp} details our simulation environment and Section~\ref{attack 
Effectiveness} evaluates the impact of interest flooding attack in our setup. Section~\ref{sec:reactive} 
presents our countermeasure, evaluated in Section~\ref{sec:evaluation}. 
Finally, Section~\ref{relatedwork} overviews related work and Section~\ref{sec:conclusion}
concludes the paper.

\section{NDN Overview}
\label{sec:overview}

NDN supports two types of messages: {\em interests} and 
{\em content}~\cite{ccnx-protocol}. 
A content message includes a name, a payload and a digital signature 
computed by the content producer.
Names are composed of one or 
more components, which have a hierarchical structure. In NDN notation, ``\ndnname{/}'' 
separates name components, e.g., \ndnname{/cnn/politics /frontpage}.
Content is delivered to consumers only upon explicit request. Each request corresponds to 
an {\em interest} message. Unlike content, interests are not signed. An interest message 
includes a name of requested content.
In case of multiple content under a given name, optional 
control information can be carried within the interest to restrict desired content. 
Content signatures provide data origin authentication. 

NDN routers forward interests towards the content producer responsible for the requested 
name, using name prefixes 
for routing. 
Each NDN router maintains a Pending Interest Table (PIT) -- a lookup table containing 
outstanding [{\em interest},{\em arrival-interfaces}] entries.
When an NDN router receives an interest, it first looks up its PIT to determine
whether another interest for the same name is currently outstanding. There
are three possible outcomes: 
(1) If the same name is already in the router's 
PIT and the arrival interface of the present interest is already in the set of 
{\em arrival-interfaces} of the corresponding PIT entry, the interest is discarded. 
(2) If a PIT entry for the same name exists, yet the arrival interface is new, the
router updates the PIT entry by adding a new interface to the set. 
The interest is not forwarded further.
(3) Otherwise, the router creates a new PIT entry and forwards the 
present interest.

Upon receipt of the interest, the producer injects content into the network, thus 
{\em satisfying} the interest. The requested content is then forwarded towards the 
consumer, traversing -- in reverse -- the path of the corresponding interest. Each router 
on the path deletes the PIT entry corresponding to the satisfied interest.
In addition, each router may cache a copy of forwarded content in its local Content Store 
(CS).
A router that receives an interest for already-cached content does not forward the 
interest further; it simply returns cached content and retains no state about the 
interest.

Not all interests result in content being returned. If an interest encounters either a 
router that cannot forward it further, or a content producer that has no such content, no 
error packets are generated. PIT entries for unsatisfied interests in intervening routers 
are removed after a predefined {\em expiration} time. The consumer can choose whether to regenerate the same interest after a timeout.

\section{Interest Flooding}
\label{approaches}
It is easy to see that an adversary can take advantage of CS and PIT -- two features unique to 
NDN -- to mount DoS/DDoS attacks specific to NDN. We focus on attacks that exploit 
the PIT, in particular, rapid generation of large numbers of interests that saturate the victim 
router's PIT. Once the PIT is completely full, all subsequent incoming (un-collapsible) interests are dropped. 
Flooding an NDN router with interests saturates its PIT 
if the rate of incoming interests is higher than the rate at which entries are removed 
from the PIT, either due to returning content or expiration. This is the goal of 
interest flooding attacks.

There are several analogies between well-known SYN flooding~\cite{Wang02detectingsyn} and 
interest flooding. In a SYN flooding attack, the adversary's goal is to consume resources on 
the victim host by initiating a large number of TCP connections. This requires the victim to 
keep state for each connection for a relatively long time.
The main difference between SYN flooding and interest flooding attacks is the victim: the 
primary victims of interest flooding are routers. End-hosts are secondary victims.

As observed in~\cite{ndn-dos}, there are at least three ways to mount this attack. 
The adversary can issue closely-spaced interests for:
(1) existing static content;
(2) dynamically-generated content; or
(3) non-existent content.
In the rest of this paper, we refer to interests for non existing content as {\em fake 
interests}.

In strategy (1), the adversary requests distinct content to avoid interest collapsing.  
If the flooding rate is sufficiently high, the producer (or, possibly, a router 
past the victim) will start dropping packets. This causes interests to linger in the 
victim's PIT until they expire, 
Depending on the victim's ability to satisfy interests 
quickly (and flush them from the PIT), this strategy may be very expensive.
Also, router caches might lower the impact of this attack, satisfying 
adversary's requests {\em before} they reach the victim. 

Strategy (2) is similar to (1), except that content is never returned from caches. Also, 
(2) may impose more load on the producer, due to the increased number of requests 
for content that can not be precomputed. This could cause higher round-trip latency 
and higher rate of dropped packets, which forces adversary's interests to 
remain in the victim's PIT longer.

Strategy (3) allows the adversary to create entries in the victim's PIT 
for which no content will ever be returned. This has several consequences: i) Adversarial interests referring to non-existent content 
are stored in the victim's PIT until they expire.
ii) The maximum rate of adversarial interests does not depend on the 
bandwidth allocated by the victim to content packets, or on the adversary's 
ability to receive content.
iii) Adversarial interests cannot be satisfied by router caches, since they 
request non-existing content.
iv) If constructed properly (for example, a with random component 
at the end of each name) adversarial interests are never collapsed. 
These effects allow the adversary to efficiently fill up the victim router's PIT, 
which makes this attack more dangerous than (1) or (2).
Therefore, in the rest of this paper, we  focus on  (3) -- interest flooding via
fake interests.

It is straightforward to construct interests that are routed through the victim. 
Let $R$ be the router advertising namespace \ndnname{/nsf/fia/}. If the adversary 
issues interests for \linebreak \ndnname{/nsf/fia/$rnd$} (where ``$rnd$'' is a random string),
they are forwarded through $R$.

\section{Evaluation Environment}
\label{exp}
We use simulations to quantify effects of both attacks and countermeasures.
In particular, we run CCNx over NS-3 \cite{ns3} via DCE.
CCNx~\cite{CCNx} is the official implementation of NDN, originally developed by 
PARC, and released as open-source project in 2009. 
Even though CCNx codebase is still 
in early stage of development, it provides all basic functionalities of NDN. 
CCNx currently runs as an overlay on top of IP. 
Direct Code Execution~\cite{DCE} (DCE) is a framework developed by INRIA to 
allow regular applications to access a network environment simulated using NS-3. 
DCE allows us to test the latest CCNx, without reimplementing it for NS-3.

\begin{figure*}[hp!]
 \centering
   {\includegraphics[width=15.5cm]{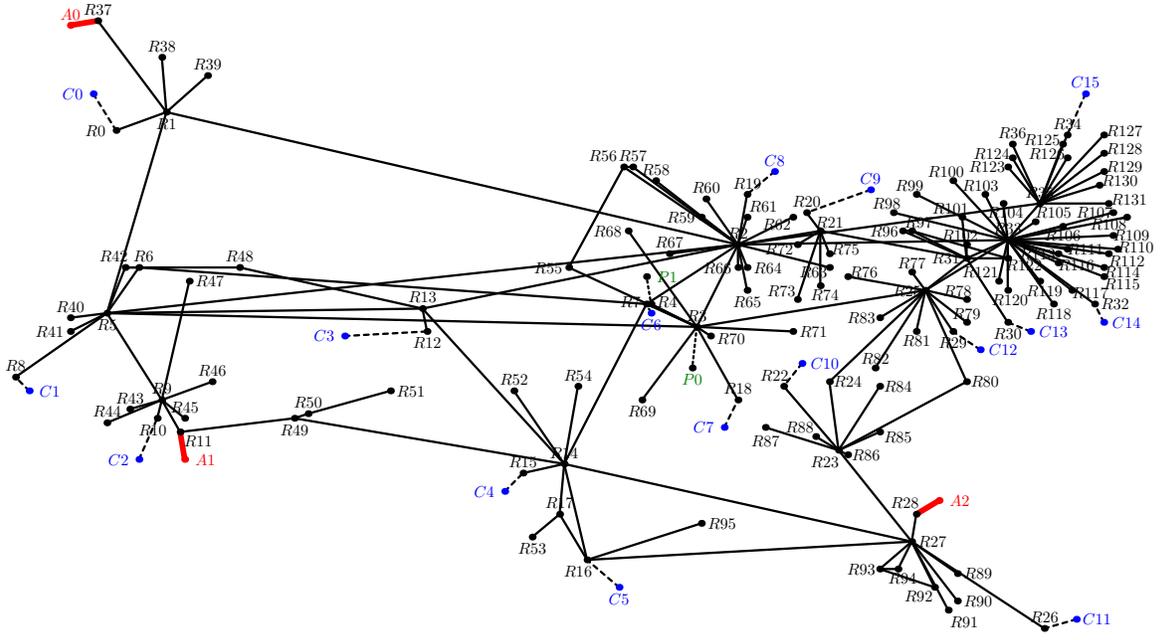}
   \label{arc:ATT}}
 \caption{AT\&T topology\label{fig:architectures}}
 \end{figure*}

We emphasize that running simulations of NDN as an overlay (over IP) reflects the status of 
the current CCNx implementation. In fact, even in the official NDN testbed~\cite{NDN},
links between routers are essentially Generic Routing Encapsulation (GRE) tunnels carrying 
UDP packets. 
Our experiments are performed over the AT\&T network topology shown in 
Figure~\ref{fig:architectures}.
We use R$x$, C$x$, P$x$, and A$x$ to denote the $x$-th 
router, consumer, producer, and adversary-controlled node, respectively. 
Continuous lines in Figure \ref{fig:architectures} indicate connections between routers; 
dashed lines denote connections between consumers and routers; dotted lines 
represent connections between producers and routers.
Our setup includes 16 (honest) consumers and 2 producers.

We first analyze the topology without any adversarial traffic. This provides us with a baseline. 
Each consumer issues interests for content produced by P0 and P1. Interests 
retrieve distinct pieces of non-existent content; therefore, routers cannot collapse them 
or satisfy them via cached content.
Consumers send a short burst of 30 interests, spaced by 2~ms, at time 
$t=1$~s of the simulation.
Starting from $t= 1.2$~s, consumers switch to a rate of one interest every $10.7$~ms. In 
our configurations, such interest spacing allows routers to forward interests roughly at 
the same rate at which they receive content packets.
We set routers' PIT size to $120$~KB, while the interest expiration time was set to the default 
timeout of $4$~s.

\begin{figure*}[]
 \centering
    \subfigure[Throughput (abs. values)]
   {\includegraphics[width=7.3cm]{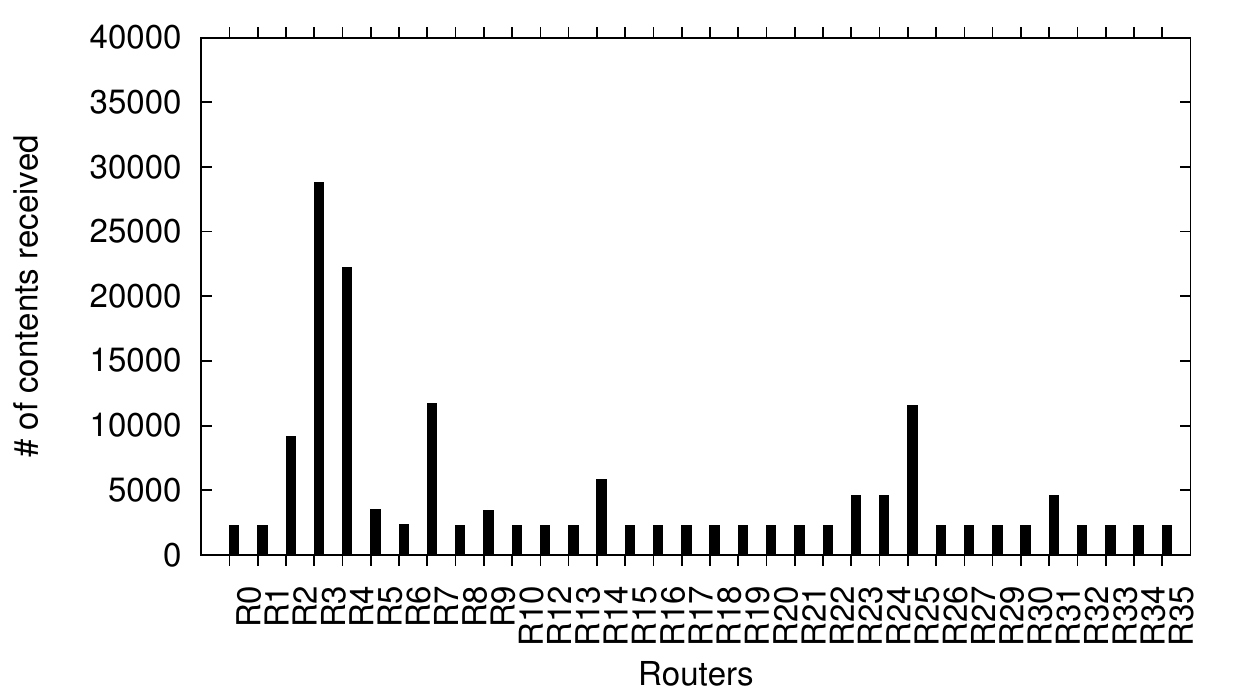}\label{fig:throughput-ATT}}
 \hspace{-3mm}
    \subfigure[PIT usage]
   {\includegraphics[width=7.3cm]{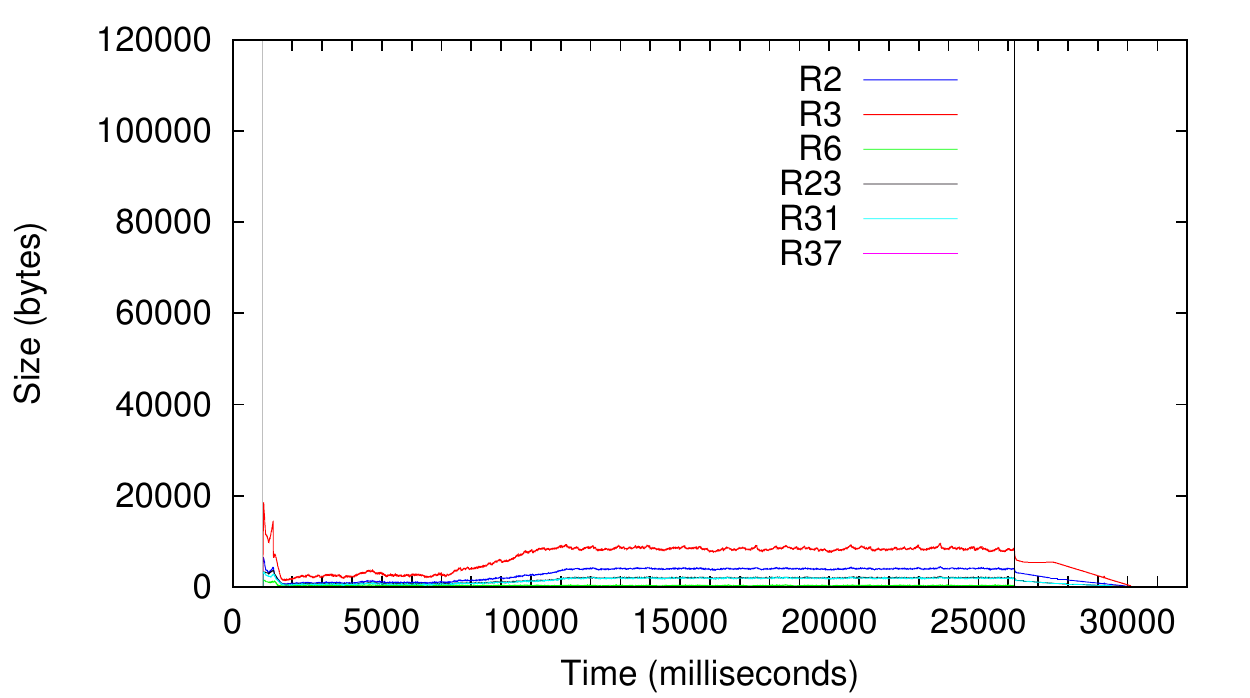}\label{fig:PIT-ATT}}
 \caption{Baseline behavior (no attack)\label{fig:throughput-PIT-baseline}}
 \end{figure*}

We report the average of the various runs in Figure \ref{fig:throughput-PIT-baseline}.
In particular, 
Figure \ref{fig:throughput-ATT} 
shows the total number of contents ($y$-axis) received by the different routers ($x$-axis), while 
Figure %
\ref{fig:PIT-ATT} shows PIT usage ($y$-axis) as a 
function of simulation time ($x$-axis). 
The maximum value on the y-axis for 
\ref{fig:PIT-ATT} 
corresponds to the total space available in the PITs ($120$~KB). Also, the two vertical 
lines (at $1$~s and $26$~s) indicate the instant when consumers start and stop 
sending interests. (The same notation is used in all graphs that refer to PIT usage 
reported in this paper.)

\section{Attack Effectiveness}
\label{attack Effectiveness}
We assume that the adversary is able to corrupt %
a portion of the consumers, through which it implements the attack -- i.e.,
issues fake interests. However the adversary is not allowed to control routers. We 
believe that this restriction is realistic and well represents the current scenario of 
DDoS attacks (e.g.,~\cite{ioncannon}).
While we do not exclude that attacks might come from internal routers, we 
leave the investigation of this as future work.

In our simulations we observe that successful instantiation of interest flooding requires very small 
amount of  bandwidth.
The adversary controls the nodes connected through a red solid line in Figure 
\ref{fig:architectures}.
The three adversarial nodes (A0, A1, A2) send interests for non-existent content for the 
namespace registered by P0 -- i.e., all fake interests are routed to P0. Similar to 
honest nodes, the adversary starts sending interests at $t=1$~s. Fake interests are 
generated every $1.337$~ms. Behavior of honest consumers is unchanged from the base scenario.

\begin{figure*}[htp!]
 \centering
 \subfigure[Throughput (\%)]
   {\includegraphics[width=7.3cm]{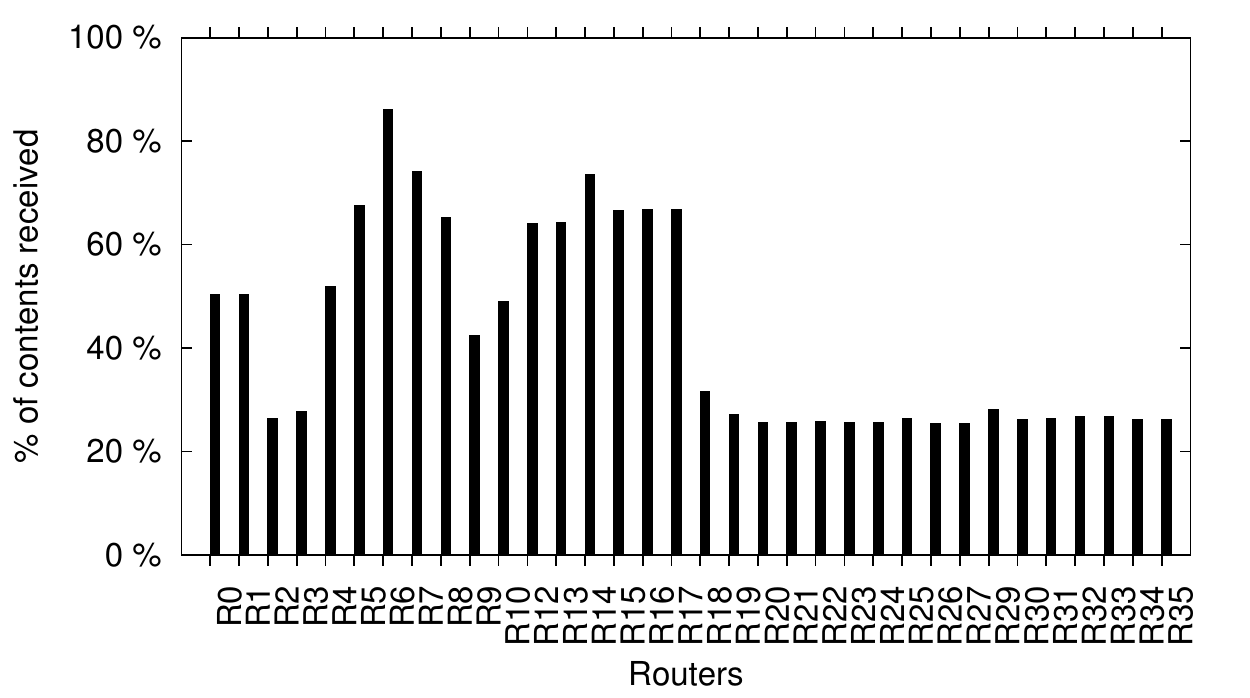}\label{fig:attack-throughput:ATT}}
 \hspace{-3mm}
   \subfigure[PIT usage]
   {\includegraphics[width=7.3cm]{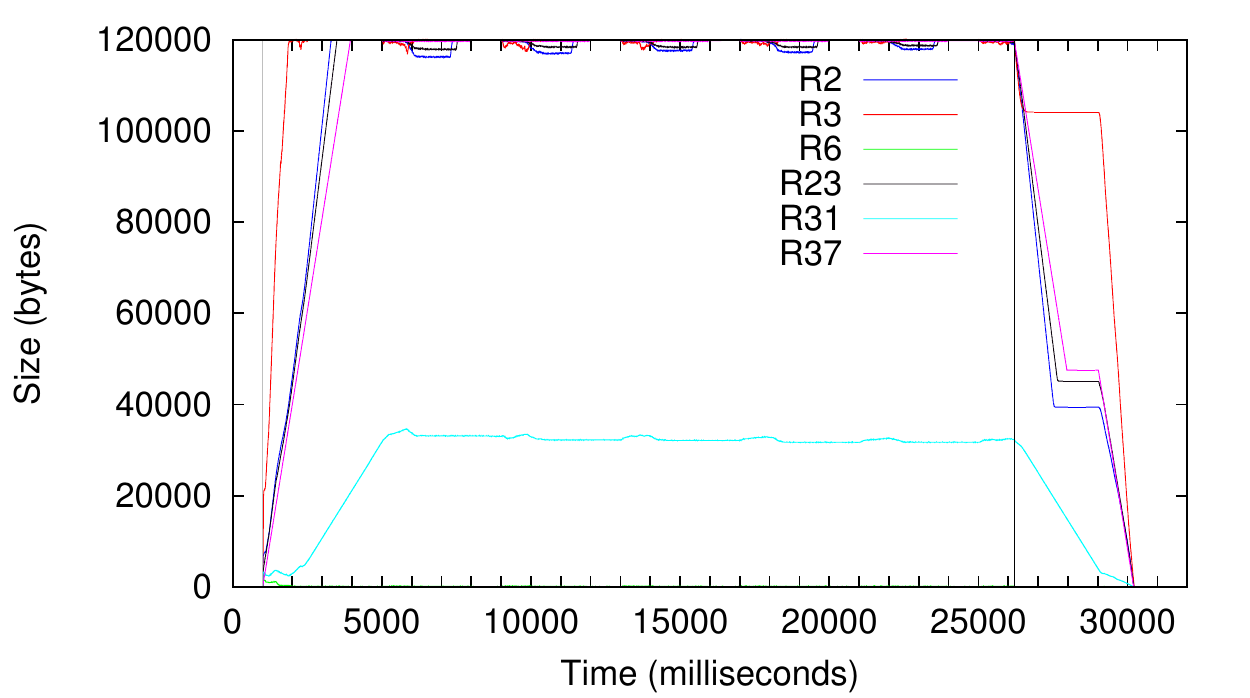}\label{fig:attack-PIT:ATT}}
 \caption{Interest Flooding Attack (IFA): impact over baseline}
\label{fig:throughput-attack}
 \end{figure*}

Attack results are plotted in Figure \ref{fig:throughput-attack} for some representative 
nodes in both topologies. In particular, 
Figure \ref{fig:attack-throughput:ATT} shows the ratio of content packets forwarded during 
the attack with respect to the same network with no malicious traffic. 
Figure \ref{fig:attack-PIT:ATT} shows PIT usage.

Figure \ref{fig:attack-throughput:ATT} demonstrates that the attack has significant impact on the network:  
several routers forward 20\% of the original traffic. 

It is important to note that 
consumers are spread all 
over the network and the number of adversaries is quite small (three for both topologies). However, 
attack impact is significant: fraction of packets forwarded by routers varies between 25\% and 80\% 
with respect to the base-line scenario. Differences in the effectiveness of the attack for different routers  
can be explained by their distance from the adversary-producer paths.
We emphasize that reduced bandwidth available to consumers can only be attributed 
to high PIT usage, as shown in 
Figure~\ref{fig:attack-PIT:ATT}.
It is easy to see that the fraction of traffic forwarded by routers drops significantly 
once PITs fill up.
As confirmed by Figure \ref{fig:attack-PIT:ATT}, R3 (closest to P0) is the first to succumb (reaching its PIT limit). This can be attributed to the central role R3 occupies in the topology.

\section{Our Countermeasure: Poseidon}
\label{sec:reactive}
We now discuss countermeasures for IFA. We focus on countermeasures that consist in a detection and a reaction phase. Detection can be local or distributed (collaborative). In the former, routers 
rely only on local metrics (e.g., PIT usage, rate of unsatisfied interests, amount of 
bandwidth used to forward content) to identify an attack. In the latter, nearby 
routers collaborate to determine whether an attack is in progress and how to mitigate it. 

In case of successful interest flooding attack, the victim router can easily identify an attack by observing 
whether its PIT is full or whether the bandwidth allocated to forwarding of content is 
very small. However, it may not be possible for a router on the path to the victim to 
detect an attack in progress. 
Collaborative detection mechanisms allow routers to exchange information about their 
state, with the goal of detecting an attack in progress as soon -- and as close to the adversary -- as possible.
With collaborative detection, routers not only exchange 
information about the {\em existence} of an attack, but also the (locally detected) 
properties of such attack: strategies can take into account feedback from multiple routers.

In this paper, we consider a particular  collaborative approach -- known as 
push-back~\cite{ndn-dos} -- to counter interest flooding. 
We call our implementation {\em Poseidon}, and we discuss it in the rest of the paper.
Poseidon is a set of algorithms that run on routers, with the goal of identifying traffic 
anomalies (especially,  interest flooding) and mitigate their effects.
Poseidon continuously monitors per-interface rates of unsatisfied interests with respect 
to overall traffic. If these rates change significantly between two consecutive time intervals, it 
sets a filter on the offending interface(s) (which reduces the number of incoming 
interests). Additionally, Poseidon can issue a push-back ``alert'' message to the 
same interfaces, to signal that an interest flooding attack is in progress.

Poseidon keeps several statistics on expired interests. In particular, for each of them it 
records namespace and incoming/outgoing interfaces information. 
Relatively common network phenomenon (e.g., packet loss) and regular applications behavior 
usually account for only a (relatively) small amount of expiring interests in routers' PITs.

In the next sections we introduce the detection and reaction phases of Poseidon. 
Notation used is shown in Table~\ref{tab:notation}.

\begin{table}[h]
\begin{center} 
\resizebox{0.44\textwidth}{!} {
\begin{tabular}{|r|l|}
\hline
${\sf R}$ &  set of all routers in the network running Poseidon
\\ \hline
$r_i$ &  $i$-th router, 1 $\le$ $i$ $\le$ $|{\sf R}|$ 
\\ \hline
$r_i^j$&	 $j$-th interface on router $r_i$ 
\\ \hline
$t_k$   &	 $k$-th time interval 
\\ \hline
$\omega(r_i^j, t_k)$ & rate between incoming interest and outgoing \\content
			& for a given interface $r_i^j$
\\ \hline
$\rho(r_i^j, t_k)$   & PIT space used by interests arrived on interface\\ $r_i^j$, 
				    & measured at the end of interval $t_k$
\\ \hline
$\Omega(r_i^j)$   & interest flooding detection threshold for $\omega(r_i^j, t_k)$
\\ \hline
$\Rho(r_i^j)$  & interest flooding detection threshold for $\rho(r_i^j, t_k)$
\\ \hline
\end{tabular}
}
\end{center}
\caption{Notation.}
\label{tab:notation}

\end{table}

\subsection{Detection Phase}
\label{detectionphase}

Attacks are detected using two parameters: $\omega(r_i^j, t_k)$, and $\rho(r_i^j, t_k)$. 
The former represents the number of incoming interests divided by the number of outgoing 
content packets, observed by a router $r_i$ on its interface $r_i^j$ within time interval 
$t_k$:
\begin{equation*}
\omega(r_i^j, t_k) = \frac{(\# \mbox{ of interests from } r_i^j \mbox{ at interval } t_k)}
{(\# \mbox{ of content packets to } r_i^j \mbox{ at interval } t_k)}.
\end{equation*}
$\rho(r_i^j, t_k)$ indicates the number of bytes used to store interests in PIT, coming from interface $r_i^j$ within time interval $t_k$.

Poseidon detects an attack when both $\omega(r_i^j, t_k)$ and $\rho(r_i^j, t_k)$ exceed their respective thresholds $\Omega(r_i^j)$ and $\Rho(r_i^j)$.
The detection algorithm is executed at fixed time intervals -- typically every 60~ms -- and in the presence of particular events, (i.e., push-back messages, as detailed below).

The parameter $\omega(r_i^j, t_k)$  is a good representation of the ability of routers to 
satisfy incoming interests in a particular time interval. (This is also confirmed by our 
experiments, detailed in Section~\ref{sec:evaluation}.) 
In particular, $\omega(r_i^j, t_k)>1$ indicates that the number of content packets 
forwarded to   $r_i^j$ is smaller than the number of interests coming from the same 
interface. However, a small bursts of (either regular or non-satisfiable) interests may 
not be caused by an attack.
Hence, taking into account only $\omega(r_i^j, t_k)$ (i.e., not considering $\rho(r_i^j, t_k)$) may cause the detection algorithm to 
report a large number of false positives. 
Applying countermeasures may, in this case, produce negative effects to the 
overall network performance.

We argue that neither increasing $\Omega(r_i^j)$, nor computing $\omega(r_i^j, t_k)$ over 
longer intervals, produces the indented effects. In fact, in the 
first case the bound must be set high enough to avoid classification of short  burst of 
interests  as  attacks; however this could inevitably lead to late- or mis-detection of 
actual attacks. 
Increasing the size of the interval over which  $\Omega(r_i^j)$ is computed may reduce the 
sensitivity of Poseidon to short burst of interests. An interval length similar or longer 
than the average round-trip time of interest/content packet, in fact, may allow (part of) 
the content requested by the burst to be forwarded back, reducing $\omega(r_i^j, t_k)$ to 
a value closer to 1.
However this could significantly increase the detection time.

Instead, to improve detection accuracy (distinguishing naturally occurring burst of 
interests from attacks), Poseidon takes into account also $\rho(r_i^j, t_k)$. This value 
measures the PIT space used by interests coming from a particular interface.  
This allows Poseidon to maintain the number of false positives low -- when compared to 
considering solely $\omega(r_i^j, t_k)$ -- while allowing it to detect low-rate interest flooding. 
In a low-rate interest flooding attack the adversary limits the rate of fake interests to keep $\omega(r_i^j, 
t_k)$ below its thresholds. Monitoring the content of the PIT allows Poseidon to observe 
the {\em effects} of the attack, rather than just its {\em causes}, allowing for early detection.

To sum up, different parameters monitored by Poseidon act as weights and counterweights 
for interest flooding detection. When a router is unable to satisfy incoming interests over a relatively short period, $\rho(r_i^j, t_k)$ may exceed the detection threshold but $\omega(r_i^j, t_k)$ will not; when the router receives a short bursts of interests, $\omega(r_i^j, t_k)$ may become larger than $\Omega(r_i^j)$ but the PIT usage will likely be within normal values. To stay undetected, an adversary willing to perform interest flooding  must therefore: (1) reduce the rate at which it sends interests, which limits the effects of the attack; and/or (2) restrict the attack to short burst, which makes the attack ineffective.

Thresholds $\Omega(r_i^j)$ and $\Rho(r_i^j)$ are not constant and may change over time to accommodate different conditions of the network. As an example, push-back messages described below provide input for determining more appropriate values for these thresholds.

\subsection{Reaction Phase}
\label{reaction-phase}

Once an interest flooding attack from interface $r_i^j$  of router $r_i$ has been identified, Poseidon limits the rate of incoming interests from that interface.
The original rate is restored once all detection parameters fall again below their corresponding thresholds.

With collaborative countermeasures, once a router detects adversarial traffic from a set of interfaces it limits their rate and issues an alert message on each of them.
An alert message is an unsolicited content packet which belongs to a reserved namespace (``\ndnname{/pushback/alerts/}'' in our implementation), used to convey information about interest flooding attack in progress. 
There are two reasons for using content packets rather than interests for carrying push-back information: (1) during an attack, the PIT of the next hop connected to the offending interface may be full, and therefore the alert message may be discarded; and (2) content packets are signed, while interests are not. This allows routers to determine whether an alert message is legitimate.

Routers running Poseidon do not process alert messages as regular content: alerts are not checked against PIT content and are not forwarded any further.
The payload of an alert packet contains: the timestamp corresponding to the alert generation time; the new (reduced) rate at which offending interests will be accepted on the incoming interface; and detailed information about the attack -- such as the namespace(s) used in malicious interests. 

Router $r_i$ receiving a packet $msg$ processes it as detailed in Algorithm \ref{alg:push-back}.
A persistent interest flooding attack on router $r_i$ causes it to send multiple alert messages towards the source(s) of the attack. Such sources will decrease their thresholds $\Omega(r_i^j)$ and $\Rho(r_i^j)$ until they detect the attack and implement rate-limiting on the malicious interests.
If no attack is reported for a predefined amount of time, thresholds are restored to their original values.

This push-back mechanism allows routers that are not the target of the attack, but are unwittingly forwarding malicious interests, to detect interest flooding early. In particular, alert messages allow routers to detect interest flooding even when they are far away from the intended victim -- i.e., close to nodes controlled by the adversary, where countermeasures are most effective.

\begin{algorithm}[ht!]
\caption{$\sf MessageProcessing$}\label{alg:push-back}
\SetKwInOut{Input}{input}\SetKwInOut{Output}{output}
\Input{Incoming packet $msg$ from $r_i^j$; $wait\_time$; $\Omega(r_i^j)$; $\Rho(r_i^j)$; 
Scaling factor $s$; Alert message $m$ from interface $r_i^j$}
\begin{algorithmic}[1]
\IF { $msg$ is ContentObject}
	\STATE $\sf process$ $msg$ as ContentObject and return
\ENDIF
\IF { $msg$ is AlertMessage}
	\IF { ${\sf Verify}(msg.signature)$ \AND ${\sf IsFresh}(msg)$ \AND \\time from last Alert received from  $r_i^j >$ \textit{wait\_time}}
	\STATE {\tt//  Push-back reaction}
			\STATE ${\sf Decrease}(\Omega(r_i^j),s)$
			\STATE ${\sf Decrease}(\Rho(r_i^j),s)$
	\ELSE
		\STATE $\sf drop$ $msg$ and return
	\ENDIF
\ENDIF
\IF{ $msg$ is Interest }
	\IF{{$\omega(r_i^j, t_k) > \Omega(r_i^j)\ \ $ \AND $\rho(r_i^j, t_k) > \Rho(r_i^j) \ $}}  %
		\STATE $\sf drop$ $msg$
		\IF {time from last Alert sent on interface $r_i^j >$ \textit{wait\_time} } 
			\STATE $\sf send$ Alert to $r_i^j$ \ENDIF
		\ELSE
		\STATE $\sf process$ $msg$ as Interest
	\ENDIF 
\ENDIF
\end{algorithmic}
\end{algorithm}

%

%
\section{Evaluation}
\label{sec:evaluation}

In this section we report on experimental evaluation of countermeasures presented in Section~\ref{sec:reactive}.
Our countermeasures are tested over the same topology used in previous experiments and detailed in Figure~\ref{fig:architectures}. Each router implements detection techniques discussed in Section~\ref{detectionphase} and countermeasures from Section~\ref{reaction-phase}.
As for the parameters used in our experiments, we considered these initial values: for each router $r_i$, interface $r_i^j$, $\Omega(r_i^j)=3$ and $\Rho(r_i^j)=1/8$ of the PIT size. 
Furthermore, we set scaling factor $s=2$ and $wait\_time$ to $60$~ms. 
The $\sf Decrease$ function divides the threshold in input by $s$ at each invocation. Similarly, the $\sf Increase$ function increases its input by 1/8 of the current value.
Consumers request the same content at the same rate as in the previous simulations. Similarly, the nodes controlled by the adversary implement interest flooding as in the simulation in Section~\ref{attack Effectiveness}.

\begin{figure*}[htp!]
 \centering
\subfigure[Rate-based]
   {\includegraphics[width=7.3cm]{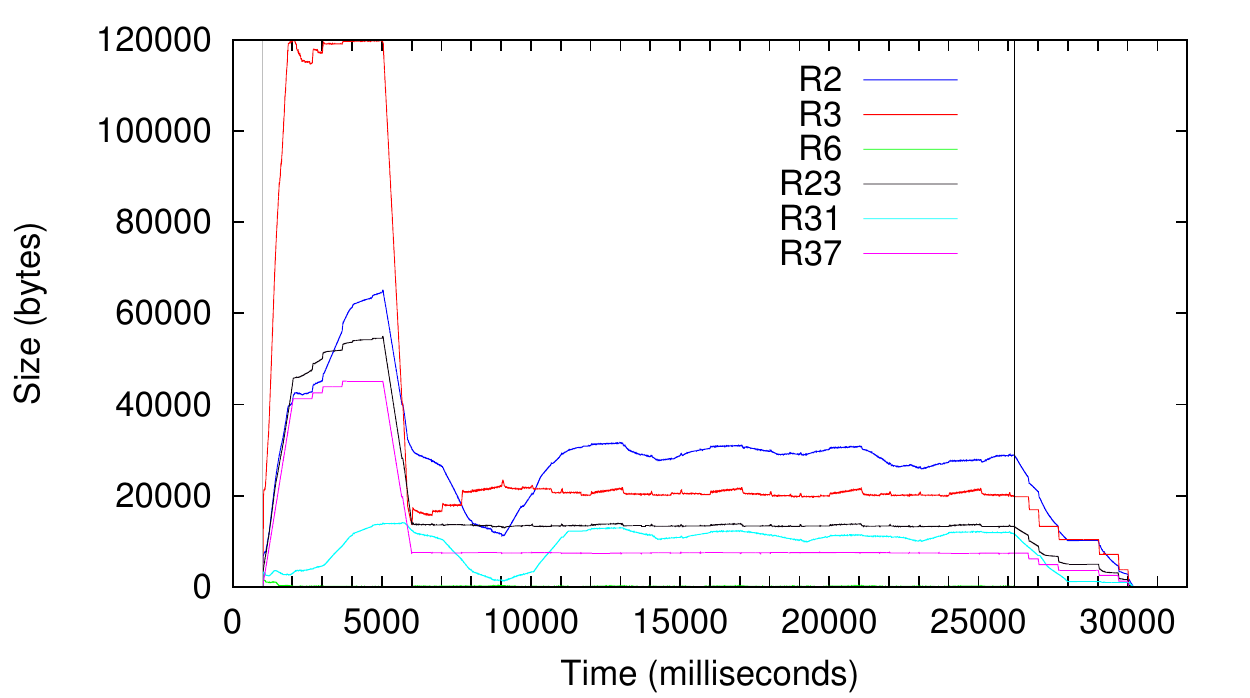}\label{rate-PIT-ATT}
\label{fig:rate-based-count}}
 \hspace{-3mm}
\subfigure[Push-back]
   {\includegraphics[width=7.3cm]{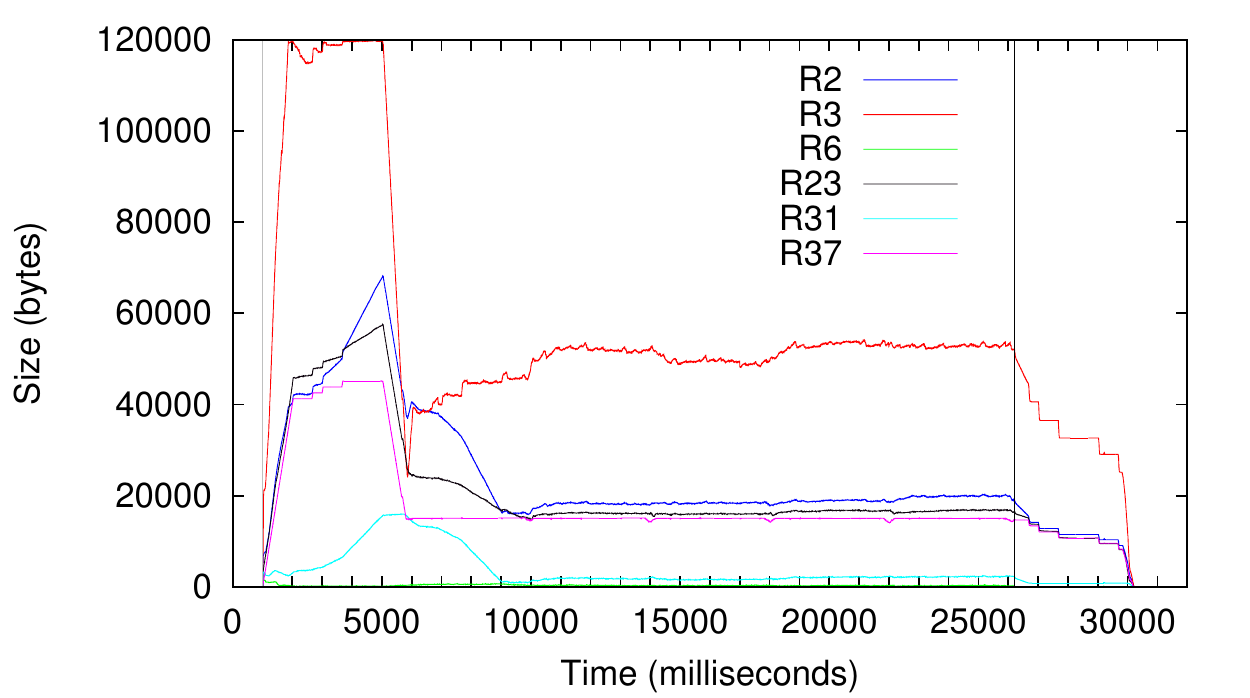}\label{push-PIT-ATT}}
 \caption{PIT usage with countermeasures}
\label{fig:push-back-count}
 \end{figure*}

\begin{figure*}[htp!]
 \centering
    {\includegraphics[width=15.7cm]{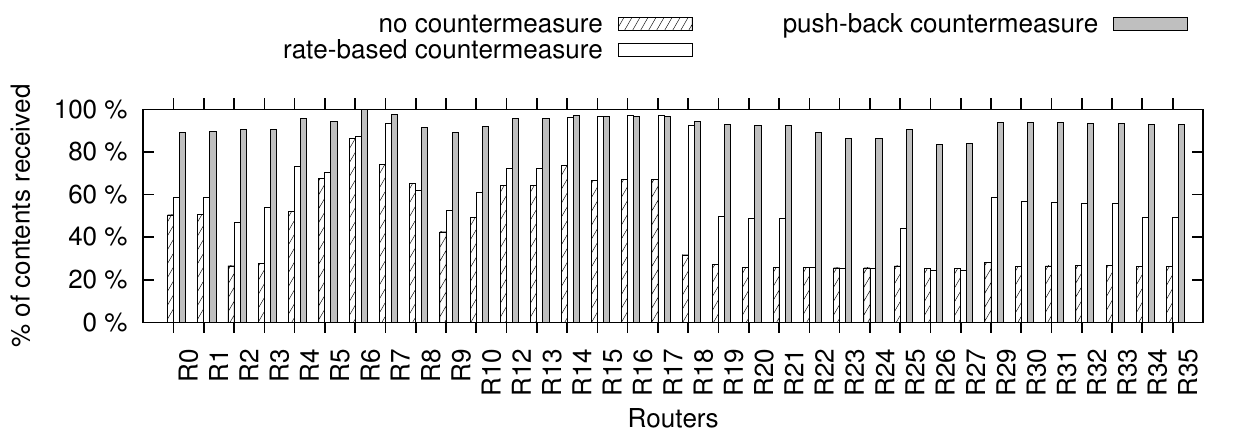}} 
 \caption{Push-back: relative throughput (\%)}
\label{push-throughput-ATT}
\label{fig:push-back-count}
 \end{figure*}

\paragraph{Local Countermeasures.} Figure \ref{fig:rate-based-count} shows the result of local countermeasures. Values shown represent the average of 20 executions.
Figure \ref{push-throughput-ATT} reports the ratio of content packets received with respect to the scenario with no adversary. (For comparison purposes, in the same figure we also report the corresponding value with no countermeasures in place.)

Our results show that the rate-based (local) countermeasure -- while simple -- is very effective (see Figure \ref{push-throughput-ATT}): under attack, the performances with the countermeasures increases by some 50\%  for most routers (e.g., see R30), when compared to the situation with the attack and no countermeasures.
The impact of the adversary  is now more limited: the attack only reduces the traffic by {\em some} 50\% for most. In contrast, without any countermeasure the adversary was able to reduce content traffic by about 80\%.
Figure~\ref{rate-PIT-ATT} report PIT usage over the same experiments, for some representative routers.
Our results also show that this countermeasure significantly reduces the PIT usage in presence of an adversary. The effects of the rate-based approach are evident at $t=6$~s, when fake interests corresponding to the initial phase of the attack -- those that triggered the detection  -- expire. This shows that the detection time for the attack is around one second. 
\paragraph{Distributed Countermeasures.} 
Figure \ref{push-throughput-ATT} show the ratio of content packets received under attack with the the push-back countermeasure in place. 
To simplify comparison, we  report the results of the simulations of the attack without countermeasures and with the previous (local) countermeasure.
The push-back mechanism offers visibly better performance compared to the rate-based countermeasure: 
for several routers the improvement with respect to the local countermeasure is over 300\%.

A similar conclusion applies also to the PIT usage -- which is another measure for attach effectiveness.
In fact, 
it is possible to observe a significant benefits of push-back comparing Figure~\ref{rate-PIT-ATT} to Figure~\ref{push-PIT-ATT}, e.g. for the PIT of router R31.

\begin{figure*}[htp!]
 \centering
 \subfigure[R4: baseline]
   {\hspace{-2mm}\includegraphics[width=5.8cm]{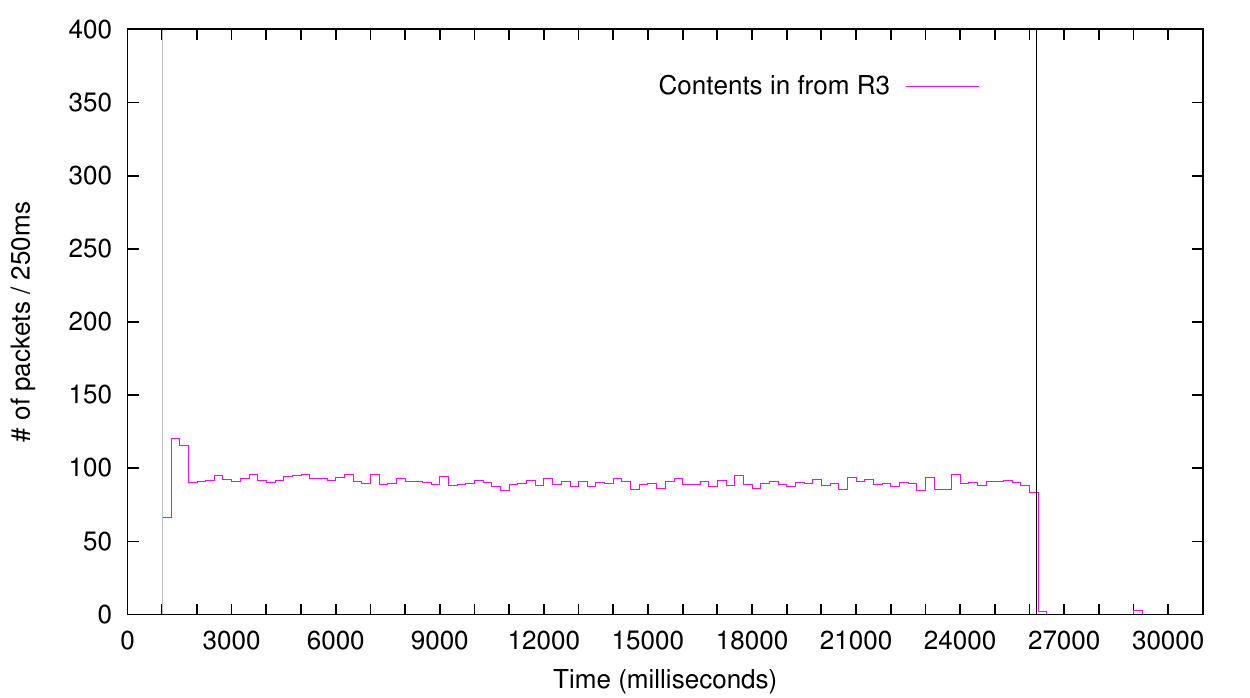}}\hspace{1mm}
 \subfigure[R4: attack]
   {\includegraphics[width=5.8cm]{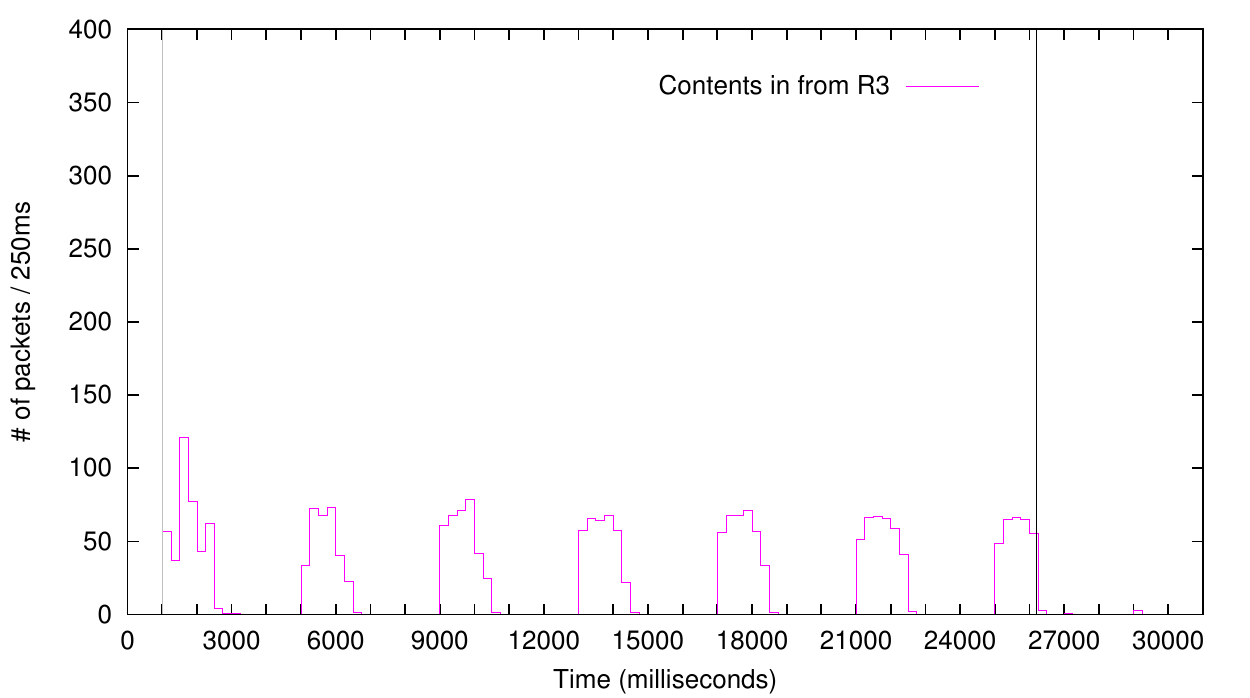}\label{ATT-R4-attack}}\hspace{1mm}
 \subfigure[R4: push-back countermeasure]
   {\includegraphics[width=5.8cm]{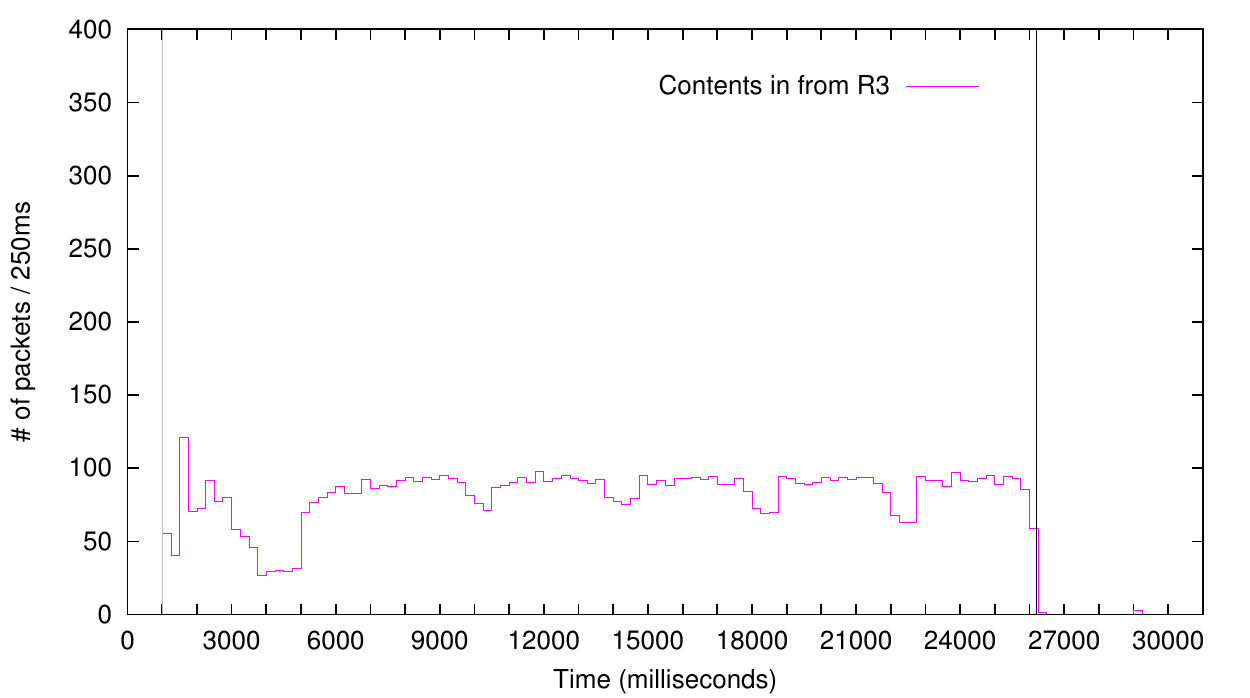}}
   \caption{AT\&T: content throughput  (abs.~values)}
\label{fig:ATT-routers-selection}
 \end{figure*}

So far we have considered cumulative results for throughput; however, it is interesting to analyze also how the content packets throughput varies over time in different scenarios.
{Figure \ref {fig:ATT-routers-selection} shows the effect of our push-back countermeasure on a router (R4), which we deem notable in our topology.
This figure clearly illustrates that using the distributed (push-back) countermeasure, routers are able to provide roughly the same throughput measured without interest flooding.}

An interesting phenomenon highlighted by 
Figure \ref{ATT-R4-attack} is the cyclical behavior of the amount of bandwidth available to content.
This pattern can be explained as follows. As soon as the PITs of these routers are filled up with fake interests, no legitimate interests are forwarded and therefore no content is routed back. After four seconds -- i.e., in our setting, the lifetime of an unsatisfied interest  -- fake interests are removed from the PITs allowing routers to forward new (legitimate) requests for content. When this happens, the adversary is quickly able to fill up PITs again. This process continues indefinitely for the whole duration of the attack.
%

%

\section{Related Work}
\label{relatedwork}

There is lots of previous work on DoS/DDoS attacks on the current Internet  
infrastructure. Current literature addresses both attacks
and countermeasures on the routing infrastructure~\cite{routing-dos},
packet flooding~\cite{packet-flooding}, reflection
attacks~\cite{reflection-attacks}, DNS cache
poisoning~\cite{And05perilsof} and SYN flooding
attacks~\cite{Wang02detectingsyn}. Proposed 
countermeasures are based on various
strategies and heuristics, including: anomaly
detection~\cite{Carl:2006:DAT:1110639.1110699},  
packet 
filtering~\cite{Tupakula:2003:PMC:783106.783137}, IP trace
back~\cite{Lu:2008:GMP:1368310.1368337,Stone:2000:CIO:1251306.1251321},
ISP collaborative defenses \cite{DBLP:journals/tpds/ChenHK07} and user-collaborative defenses~\cite{GkantsidisR06}.
The authors of~\cite{ndn-dos} present a spectrum of possible DoS/DDoS attacks in NDN. They classify those attacks in interest flooding and content/cache poisoning, and provide a high-level overview of possible countermeasures. However, the paper does not analyze specific attacks or evaluate countermeasures. 

NDN caching performance optimization has been recently investigated 
with respect  to various metrics including energy
impact~\cite{VanSmBriPlStThBra09-Voice,approximate,greening}.
The work of Xie, et
al.~\cite{enhancing-cache-robustness}
address cache
robustness in NDN. This work introduces CacheShield, a proactive
mechanism that helps routers to prevent caching unpopular content and
therefore maximizing the use of cache for popular one.
To address the same attack, Conti et al~\cite{ContiGT13} introduce a lightweight reactive mechanism 
for detecting cache pollution attacks. 

Afanasyev et al.~independently address interest flooding in~\cite{ErsinIFIPNetworking}.
Their work confirms the feasibility of interest flooding attack, and the need for an effective 
countermeasure. Interestingly, this work can be considered complementary to ours, both in terms of 
attack evaluation and countermeasures. In fact, while our experiments rely on the official NDN 
implementation \cite{CCNx}, the work in \cite{ErsinIFIPNetworking} used NDNsim \cite{uclasim}.
Although both approaches provide valuable insights into the attack, we argue that using the actual 
NDN code may result in a more accurate assessment.

A slightly different approach has been proposed by Dai et al.~in~\cite{NOMENDai}. Their technique 
relies on the collaboration between routers and producers in charge of the namespaces to which fake interests are directed. %

In \cite{WahlishSV12}, W\"ahlisch et al.~independently investigate how data-driven state can be 
used to implement various DoS/DDoS attacks. Relevant to our work, their analysis 
includes: 
resource exhaustion, which is analogous to our interest flooding attack;  
mobile blockade, in which a wireless node issues a large number of interests and then disconnects 
from the network, causing the returned content to consume a large portion of the shared network 
bandwidth;
and state decorrelation attacks, in which an adversary issues updates of local content or cache 
appearances at a frequency that exceeds the content request routing convergence. 
Attacks are tested on two physical (i.e., not simulated) topologies 
comprised of three and five NDN routers.

\section{Conclusion}
\label{sec:conclusion}
In this paper we discussed interest flooding-based DDoS over NDN. We provided, to the best of our knowledge, the first experimental evaluation of the attack. Our experiments are based on the official NDN implementation codebase; we argue that this setup provides reliable results, and closely mimics the behavior of physical (non-simulated) networks. 

We demonstrated that interest flooding attack is a realistic threat; in particular, we showed that an adversary with limited resources can reduce the amount of  bandwidth allocated for content objects to 15-25\% of the total bandwidth.
We then introduced Poseidon, a new mechanism for detecting and mitigating interest flooding. 
Poseidon relies on both local metrics and collaborative techniques for early detection of interest flooding.
We showed that the benefits of Poseidon are significant: in fact, most routers running our countermeasure are able to use more than 80\% of the available bandwidth during the attack.
\vspace{0.5cm}

\balance
\bibliographystyle{abbrv}
{
\bibliography{references}
}

\end{document}